\documentclass{IEEEtran}

\usepackage{amsmath,amssymb,amsfonts}
\usepackage{graphicx}
\usepackage{booktabs}
\usepackage{array}
\usepackage{caption}
\usepackage{subcaption}
\usepackage{placeins}
\usepackage{cite}
\usepackage{url}
\usepackage{textcomp}
\usepackage{comment}
\usepackage{float}

\begin{document}

\title{A Context-aware Gated Convex Mixtures of LSTM Experts for Nonlinear System Identification}

\author{%
Ali~Rajabi, \textit{Graduate Student Member, IEEE},
Iman~Salehi, \textit{Member, IEEE}}

\maketitle

\begin{abstract}
This work addresses nonlinear and nonstationary system identification using
one-step-ahead prediction on the nonlinear autoregressive moving average
benchmark with ten-step memory (NARMA-10). Baseline models, including
autoregressive models with exogenous input (ARX), nonlinear ARX using a
multilayer perceptron (NARX--MLP), and a single long short-term memory
network (LSTM), are used to contextualize prediction performance. To improve
robustness, multiple LSTM experts are combined through a convex mixture. A
standard adaptive convex mixture updates the mixing weights using an
error-driven rule with simplex projection, but this mechanism is reactive,
hand-tuned, and not end-to-end learnable. The proposed method introduces a context-aware
end-to-end mixture-of-experts (MoE) framework in which a differentiable
softmax gating network learns context-aware mixing weights jointly with the
expert parameters. On stationary NARMA-10, the proposed MoE gating approach
achieves performance comparable to the adaptive convex mixture. Under
regime-switching dynamics, a controlled frozen-experts ablation isolates the
mixing-weight update mechanism and shows that the MoE gate significantly
improves robustness, achieving approximately fivefold lower overall test
error and about two-and-a-half-fold lower after-switch error.
\end{abstract}

\begin{IEEEkeywords}
System identification, NARMA-10, LSTM, Convex mixture of experts,
Mixture of experts gating network.
\end{IEEEkeywords}

\section{Introduction}
\IEEEPARstart{S}{ystem} identification develops predictive models of
dynamical systems from measured input--output data and underpins monitoring,
forecasting, and control design~\cite{ljung1998system}. In modern applications,
accurate predictors are increasingly needed for plants that are nonlinear
and whose operating conditions vary over time. Such variability arises in
many settings, such as changing loads, component aging, or mode switching,
and it can degrade models that assume fixed dynamics.

A standard objective in data-driven identification is one-step-ahead
prediction: given a finite history of the measured output and input,
predict the next output and measure accuracy using a squared-error
criterion. Classical linear structures, including ARX, are introduced because
they are simple and interpretable~\cite{ljung1998system}. However, benchmarks
such as NARMA-10 contain multiplicative nonlinearities and delayed input
interactions, so purely linear predictors can underfit and yield large
error for nonlinear systems~\cite{billings2013nonlinear}.

A common solution is to keep the same history representation while increasing
model expressiveness. NARX models learn nonlinear mappings
from history features and can improve one-step accuracy on nonlinear
plants~\cite{billings2013nonlinear}. NARX implementations are widely
used because MLPs can approximate complex
nonlinear functions~\cite{lin1996learning}. Nevertheless, NARX predictors still
rely on a fixed feature vector and do not maintain an explicit internal
state~\cite{billings2013nonlinear}. As a result, the effective memory of the model is determined by the hand-tuned fixed lag orders, and the predictor must infer all temporal context from the same fixed feature representation at every time step. This becomes limiting when the true dependency horizon is uncertain, when longer histories are required, or when dynamics vary over time~\cite{lin1996learning}.

Recurrent neural networks (RNNs) address this limitation by processing a
sequence and maintaining a hidden state that summarizes relevant history. LSTM networks are a standard RNN variant that
uses gating to manage information flow and alleviate vanishing-gradient
effects~\cite{hochreiter1997long}. For nonlinear system identification, LSTMs
often provide a strong baseline because they can learn temporal
dependencies directly from windowed input--output sequences~\cite{srivastava2016lstm}. LSTM-based dynamic system identification has also been studied in papers published in American Control Conference (ACC), including Wang's convex LSTM-based identification framework
\cite{wang2017new} and a sequence-to-sequence LSTM model for piezo-electric
actuator identification \cite{yin2023sequence}.

While a single LSTM often provides a strong baseline for nonlinear system
identification~\cite{wang2017new}, its performance can be sensitive to
training stochasticity and model capacity. In particular, LSTM training is
nonconvex, and different random initializations or minibatch orders can lead
to different local solutions with noticeably different generalization error.
Moreover, a single network represents the dynamics using one set of
parameters, which may not be equally well-suited to all portions of the data
and becomes especially restrictive when the underlying dynamics change over
time. These considerations motivate using an ensemble of LSTM experts and
combining their outputs through a convex mixture, which can reduce variance,
improve robustness, and enable specialization across operating
conditions~\cite{ahmed2011variational}.

A widely used baseline is an adaptive convex mixture in which the mixing
weight vector $\alpha(k)$ is updated using an error-driven rule and
projected onto the simplex~\cite{wang2017new}. This
mechanism has two drawbacks: First, it is reactive: the mixing weights shift only
after errors increase, which can be suboptimal around regime changes in data.
Second, it is heuristic and hand-tuned: The mixing weights step size and projection
behavior must be chosen manually, and these choices can affect stability and responsiveness of the model.

Mixture models and mixtures of experts have a long history in machine
learning and adaptive systems~\cite{jacobs1991adaptive}. Hierarchical and
gating-based mixtures are also classical~\cite{jordan1994hierarchical}. For deep
learning, sparsely gated MoE models have enabled
scalable conditional computation~\cite{shazeer2017outrageously}. A survey of
mixture-of-experts formulations is given in~\cite{yuksel2012twenty}. In control
and identification, switching and regime-dependent dynamics are often
studied under Markov jump or switching-system frameworks~\cite{hamilton1989new}.
A comprehensive treatment of Markov jump linear systems is presented
in~\cite{do2005discrete}. In a related learning-to-decide setting,~\cite{amani2026learning}
formulate post-earthquake crew dispatch as a sequential decision problem
and replace classical mixed-integer programs with a transformer-based deep
reinforcement learning policy. Extending this learning-based
restoration perspective to weather-driven disruptions,~\cite{amani2026event} develop an event-driven
actor--critic dispatcher that updates crew-to-repair assignments as new
outage information and operational events are revealed.
Learning-based surrogates for classical optimization algorithms have also been explored; for example, \cite{amani2025learning} learns an interior-point method for alternating current (AC) and direct current (DC) optimal power flow. Recent ACC papers have also studied data-driven nonlinear system identification and learning-based modeling from a control-oriented
perspective. For example, Koopman representations have been used for
nonlinear throttle-valve identification \cite{bongiovanni2024data},
and recurrent neural-network models have been studied with robustness and
stability considerations \cite{revay2020convex}. These works reinforce the
growing role of learned dynamical models in control, but they do not address
the specific problem studied here: context-aware convex mixing of multiple
LSTM experts under regime-switching dynamics.

For time-varying or nonstationary systems, a single global predictor may struggle because the same parameter set must represent multiple operating
conditions. Switching and hybrid-system identification methods address this
issue by allowing the active model to change with the operating regime, but
they often require explicit mode assumptions or separate model-selection
logic~\cite{paoletti2007identification}. Mixture-of-experts models provide a softer
alternative: rather than selecting one model discretely, they combine expert
outputs through weights that can vary with context. This makes MoE models
naturally suited to regime-switching identification when different experts
capture different local dynamics.
 Most closely related to this project, Wang
introduced a convex-based LSTM concept for dynamic system identification
in the American Control Conference~\cite{wang2017new}.

This paper makes three contributions. First, it provides a complete and
reproducible evaluation pipeline on NARMA-10, including stationary and
regime-switching benchmarks and consistent error evaluation metrics. Second, it
implements and analyzes an adaptive convex mixture of LSTM experts that
updates mixing weights using per-expert batch losses and simplex
projection. Third, it proposes a fully differentiable MoE gating network
alternative that learns the mixing weights from recent input--output context using
a softmax gate trained jointly with experts. The paper also focuses on improving mixture-based LSTM predictors for settings
in which the best expert may change over time. Instead of a single
predictor, a mixture combines multiple experts using
mixing weights. The mixture can reduce variance and enable
specialization, but performance depends critically on how mixing weights
are produced and adapted. A controlled frozen-experts
ablation isolates the mixing weights update rule and demonstrates that mixing weights learned through gating network can
yield a clear advantage when expert specialization is present.

\vspace{0.5\baselineskip}
\paragraph*{\textit{Notation}}
The discrete-time index is denoted by
$k \in \mathbb{Z}_{\ge 0}$, and $u(k)\in\mathbb{R}$ and
$y(k)\in\mathbb{R}$ denote the scalar input and output signals at time
$k$. The corresponding one-step prediction is $\hat{y}(k)$. For ARX and
NARX predictors, $\varphi(k)\in\mathbb{R}^{n_y+n_u}$ denotes the history
feature vector formed from the most recent $n_y$ outputs and $n_u$ inputs.
For LSTM-based predictors, $s(k)\in\mathbb{R}^{L\times 2}$ denotes the
input window of length $L$, and $x_t$ is its $t$-th element. The LSTM hidden
and cell states are $h_t,c_t\in\mathbb{R}^{d_h}$. For mixture predictors,
$\hat{y}_i(k)$ is the prediction of expert $i$, and
$\alpha(k)=[\alpha_1(k),\ldots,\alpha_n(k)]^\top$ denotes the mixing-weight
vector. The weights satisfy $\alpha(k)\in\Delta$, where
$\Delta=\{\alpha\in\mathbb{R}^{n}:\alpha_i\geq0,\ \sum_i\alpha_i=1\}$.
The operators $\sigma(\cdot)$ and $\tanh(\cdot)$ denote the logistic
sigmoid and hyperbolic tangent, respectively, $\odot$ denotes elementwise
multiplication, and $\mathrm{MSE}$ denotes mean-squared error.
\vspace{0.5\baselineskip}

\section{Preliminaries}

\subsection{NARMA-10 Benchmark}
The NARMA-10 benchmark is a standard nonlinear autoregressive
moving-average task with approximately ten-step memory~\cite{atiya2000new}.
Given an input sequence $u(k)$, the output is generated by
\begin{equation}
\begin{aligned}
\small
y(k{+}1) ={}& 0.3\,y(k) + 0.05\,y(k)\sum_{i=0}^{9} y(k{-}i) \\
            & + 1.5\,u(k)\,u(k{-}9) + 0.1,
\end{aligned}
\label{eq:narma10}
\end{equation}
where the sum term introduces nonlinear dependence on the recent output
history and the product term introduces nonlinear dependence on delayed
inputs. Equation~\eqref{eq:narma10} is used to generate training,
validation, and test sequences under a stationary regime; regime-switching
variants are described later.

\subsection{Supervised Dataset Construction}
For one-step prediction, the learning problem is cast as supervised
regression. For ARX/NARX models, the history feature vector is
\begin{equation}
\varphi(k) = \big[y(k{-}1),\ldots,y(k{-}n_y),\,u(k{-}1),\ldots,u(k{-}n_u)\big]^{\top},
\end{equation}
where $n_y$ and $n_u$ are output and input history lengths. For
LSTM-based models, a windowed sequence input is used:
\begin{equation}
\small
s(k) = \begin{bmatrix} y(k{-}L) & u(k{-}L) \\ \vdots & \vdots \\ y(k{-}1) & u(k{-}1) \end{bmatrix} \in \mathbb{R}^{L\times 2},
\end{equation}
where $L$ is the window length. In both cases, the supervised target is
$y(k)$.

\subsection{Training Objective}
Given a predictor $\hat{y}(k)$, the mean-squared error (MSE) over a
dataset of $N$ samples is
\begin{equation}
\small
\mathrm{MSE}_{\mathrm{all}} = \frac{1}{N}\sum_{k=1}^{N}\big(y(k)-\hat{y}(k)\big)^2.
\label{eq:mse-all}
\end{equation}
For regime-switching sequences, an additional metric focuses on short
windows immediately after each regime change. Let $\mathcal{S}$ be the set
of switch indices in the test sequence and let $W$ be the window length.
The after-switch window MSE is defined as
\begin{equation}
\small
\mathrm{MSE}_{\mathrm{sw}} = \frac{1}{|\mathcal{S}|}\sum_{s\in\mathcal{S}}\!\left(\frac{1}{W}\sum_{k=s}^{s+W-1}\!\big(y(k)-\hat{y}(k)\big)^2\right)\!.
\label{eq:mse-sw}
\end{equation}
Here, $\mathrm{MSE}_{\mathrm{all}}$ measures average performance over the
entire test horizon, while $\mathrm{MSE}_{\mathrm{sw}}$ isolates transient
performance during regime transitions.

\subsection{Baseline Predictors (Context)}
The baseline suite is included to contextualize mixture performance. An
ARX predictor is linear in $\varphi(k)$:
\begin{equation}
\small
\hat{y}(k) = w^{\top}\varphi(k),
\end{equation}
where $w\in\mathbb{R}^{n_y+n_u}$ is estimated by (regularized) least
squares~\cite{ljung1998system}.

A NARX--MLP predictor replaces the linear map with a multilayer perceptron
(MLP) while keeping the same history features. In particular, the input is
$\varphi(k)$ and the output is $\hat{y}(k)$:
\begin{equation}
\small
\hat{y}(k) = \mathrm{MLP}(\varphi(k);\theta),
\label{eq:narx-mlp}
\end{equation}
where $\theta$ denotes all MLP weights and biases~\cite{lin1996learning}.
The function $\mathrm{MLP}(\varphi(k))$ consists of $H$ nonlinear hidden layers (affine map and activation) followed by a linear output layer. Because MLPs are universal function
approximators, \eqref{eq:narx-mlp} can capture nonlinearities that ARX
cannot~\cite{billings2013nonlinear}. However, NARX remains a fixed-window
regressor: it does not maintain an internal state and does not adapt its
representation as a sequence is processed.

\subsection{LSTM Cell Equations}
An LSTM processes the window $s(k)$ sequentially. Let
$x_t=[y(k{-}L{+}t{-}1),u(k{-}L{+}t{-}1)]^{\top}\in\mathbb{R}^{2}$ be the
$t$-th element of the window, and let $h_t,c_t\in\mathbb{R}^{d_h}$ denote
the hidden and cell states with hidden dimension $d_h$. The LSTM update is
\begin{subequations}
\label{eq:lstm}
\small
\begin{align}
f_t &= \sigma(W_f x_t + U_f h_{t-1} + b_f), \\
i_t &= \sigma(W_i x_t + U_i h_{t-1} + b_i), \\
g_t &= \tanh(W_g x_t + U_g h_{t-1} + b_g), \\
c_t &= f_t\odot c_{t-1} + i_t\odot g_t, \\
o_t &= \sigma(W_o x_t + U_o h_{t-1} + b_o), \\
h_t &= o_t\odot\tanh(c_t).
\end{align}
\end{subequations}
In~\eqref{eq:lstm}, $t\in\{1,\ldots,L\}$ indexes the position within the
window. The vectors $f_t,i_t,o_t\in\mathbb{R}^{d_h}$ are the forget, input,
and output gates, and $g_t\in\mathbb{R}^{d_h}$ is the candidate update.
The cell state $c_t$ stores long-term memory, while the hidden state
$h_t$ is the exposed representation passed to the next step. The matrices
$W_{\{\cdot\}}\in\mathbb{R}^{d_h\times 2}$,
$U_{\{\cdot\}}\in\mathbb{R}^{d_h\times d_h}$, and biases
$b_{\{\cdot\}}\in\mathbb{R}^{d_h}$ are trainable parameters, and
$\sigma(\cdot)$ is the logistic sigmoid~\cite{hochreiter1997long,srivastava2016lstm}.
Initial states $h_0$ and $c_0$ are set to zeros in the implementation.
After processing $L$ steps, a linear head maps the final hidden state
$h_L$ to the one-step prediction
\begin{equation}
\small
\hat{y}(k) = W_{\mathrm{out}} h_L + b_{\mathrm{out}},
\end{equation}
with $W_{\mathrm{out}}\in\mathbb{R}^{1\times d_h}$ and
$b_{\mathrm{out}}\in\mathbb{R}$.

\subsection{Convex Mixtures and the Simplex}
A convex mixture combines expert predictions using weights on the
probability simplex:
\begin{equation}
\small
\hat{y}(k) = \sum_{i=1}^{n}\alpha_i(k)\,\hat{y}_i(k),
\label{eq:convex-mix}
\end{equation}
where $\alpha(k)\in\Delta$ and
\begin{equation}
\small
\Delta = \Big\{\alpha\in\mathbb{R}^{n}:\alpha_i\geq 0,\;\sum_{i=1}^{n}\alpha_i=1\Big\}.
\end{equation}

\section{Methodology}
This section describes the two mixture mechanisms compared in this work.
Both methods use the same set of LSTM experts and the same one-step
prediction objective. They differ only in how the mixing weights are
computed. The adaptive convex mixture updates weights from recent expert
errors, whereas the proposed context-aware MoE learns a differentiable mapping from
recent input--output context to the simplex.

\subsection{Adaptive Convex Mixture of LSTM Experts}
We instantiate $n$ LSTM experts. For each time index $k$, every expert
receives the same window $s(k)$ and produces an individual one-step
prediction $\hat{y}_i(k)$. The mixture prediction is the convex combination:
\begin{equation}
\small
\hat{y}(k) = \sum_{i=1}^{n}\alpha_i(k)\,\hat{y}_i(k),
\end{equation}
with $\alpha(k)\in\Delta$.

The adaptive baseline updates $\alpha(k)$ from per-expert batch errors.
For a mini-batch of size $B$, expert $i$ incurs batch MSE:
\begin{equation}
\small
\mathrm{MSE}_i = \frac{1}{B}\sum_{b=1}^{B}\big(y^{(b)}-\hat{y}^{(b)}_i\big)^2,\quad i=1,\ldots,n,
\end{equation}
where $y^{(b)}$ is the ground-truth target and $\hat{y}^{(b)}_i$ is $i^{th}$ expert's prediction for sample $b$, and $B$ is the batch size. Let
$m=[\mathrm{MSE}_1,\ldots,\mathrm{MSE}_n]^{\top}\in\mathbb{R}^{n}$ collect
these losses. The adaptive weight update is
\begin{equation}
\small
\alpha(k)\leftarrow \Pi_{\Delta}\!\big(\alpha(k)-\eta_\alpha\,m\big),
\label{eq:adaptive}
\end{equation}
where $\eta_\alpha>0$ is the mixing weights step size and $\Pi_{\Delta}(\cdot)$
projects onto the simplex $\Delta$. We use the efficient simplex
projection algorithm in~\cite{duchi2008efficient}. Intuitively,
\eqref{eq:adaptive} decreases weights for high-error experts and
increases weights (relative) for low-error experts, while enforcing
$\alpha(k)\in\Delta$.

\subsection{MoE Gate for Mixing Weights}
The proposed method replaces the heuristic update~\eqref{eq:adaptive}
with a learned gate that outputs mixing weights as a differentiable
function of context. Let $\psi(k)\in\mathbb{R}^{d_\psi}$ denote a gate
input built from recent input--output history, such as the flattened
window $\mathrm{vec}(s(k))\in\mathbb{R}^{2L}$. The gate is a small
multilayer perceptron (MLP) that maps $\psi(k)$ to logits
$z(k)\in\mathbb{R}^{n}$:
\begin{equation}
z(k) = g(\psi(k);\theta_g),\quad \alpha(k) = \mathrm{softmax}(z(k)),
\label{eq:moe}
\end{equation}
where $\theta_g$ denotes gate parameters. The softmax ensures
$\alpha(k)\in\Delta$ automatically (nonnegative and summing to one), so no
projection step is required. Because~\eqref{eq:moe} is differentiable,
gradients of the prediction loss propagate through $\alpha(k)$ into the
gate parameters, enabling end-to-end learning of how to mix experts. This is the key distinction from the adaptive convex update. In the
adaptive method, the weights are adjusted only after expert losses are
observed and the update step is controlled by a manually selected
$\eta_\alpha$. In the MoE formulation, the gate parameters are optimized
directly with the prediction objective, so the model can learn recurring
patterns in the signal history that indicate which expert should receive
larger weight. In our WindowGate variant, $\psi(k)$ is constructed from a short window
(the same $L$ used by the experts), so the gate can detect
regime-dependent patterns in recent $y$ and $u$. This is particularly
relevant under regime switching, where the best expert may change over
time.

\subsection{Training Procedure}
All models are trained for one-step prediction using the MSE objective
\begin{equation}
\small
\mathcal{L} = \frac{1}{B}\sum_{b=1}^{B}\big(y^{(b)}-\hat{y}^{(b)}\big)^2,
\end{equation}
where $\hat{y}^{(b)}$ is the model output for sample $b$, and $B$ is the batch size. We use the
Adam optimizer~\cite{kingma2014adam} and monitor validation MSE for model
selection. For LSTM-based models, gradient clipping is applied to improve
stability (clipping threshold $1.0$ in our experiments). Hyperparameters
such as hidden size $d_h$, learning rate, and number of epochs are
selected by validation performance.

\subsection{Regime Switching and Frozen-Experts Ablation}
To stress-test adaptive convex mixture of experts mechanisms, we create regime-switching datasets by
changing the NARMA-10 coefficients periodically. Let switch times be
$s_1,s_2,\ldots,s_M$ (in samples), and let $W$ denote a short evaluation
window length after each switch.

We report two complementary test metrics. The overall test MSE averages
squared error over the entire test sequence presented in \eqref{eq:mse-all}
where N is the total number of evaluated samples in the test sequence. The after-switch window MSE focuses on adaptation immediately after
switches, which is shown in \eqref{eq:mse-sw}
where $M = |\mathcal{S}|$ is the number of regime switches, and W is the window legth after each switch. The difference is scope where \eqref{eq:mse-all} captures average
performance across all regimes and steady-state regions,
while~\eqref{eq:mse-sw} isolates transient error right after regime
changes.

Finally, we conduct a controlled apples-to-apples ablation to isolate
the mixing weights update rule. Experts are first trained on regime-specific data so
that each expert specializes, which makes the experts trained in a context-aware way. We then freeze all expert parameters and
compare (i) adaptive $\alpha(k)$ updates and (ii) the MoE gate, using
the same fixed expert outputs. Under this setting, any performance gap
is attributable to the mixing weight update mechanism rather than differences
in expert capacity.

\section{Simulation Results}

\subsection{Stationary NARMA-10 benchmark}
\begin{figure}[!h]
\centering
\begin{subfigure}{0.9\linewidth}
  \centering
  \includegraphics[width=\linewidth]{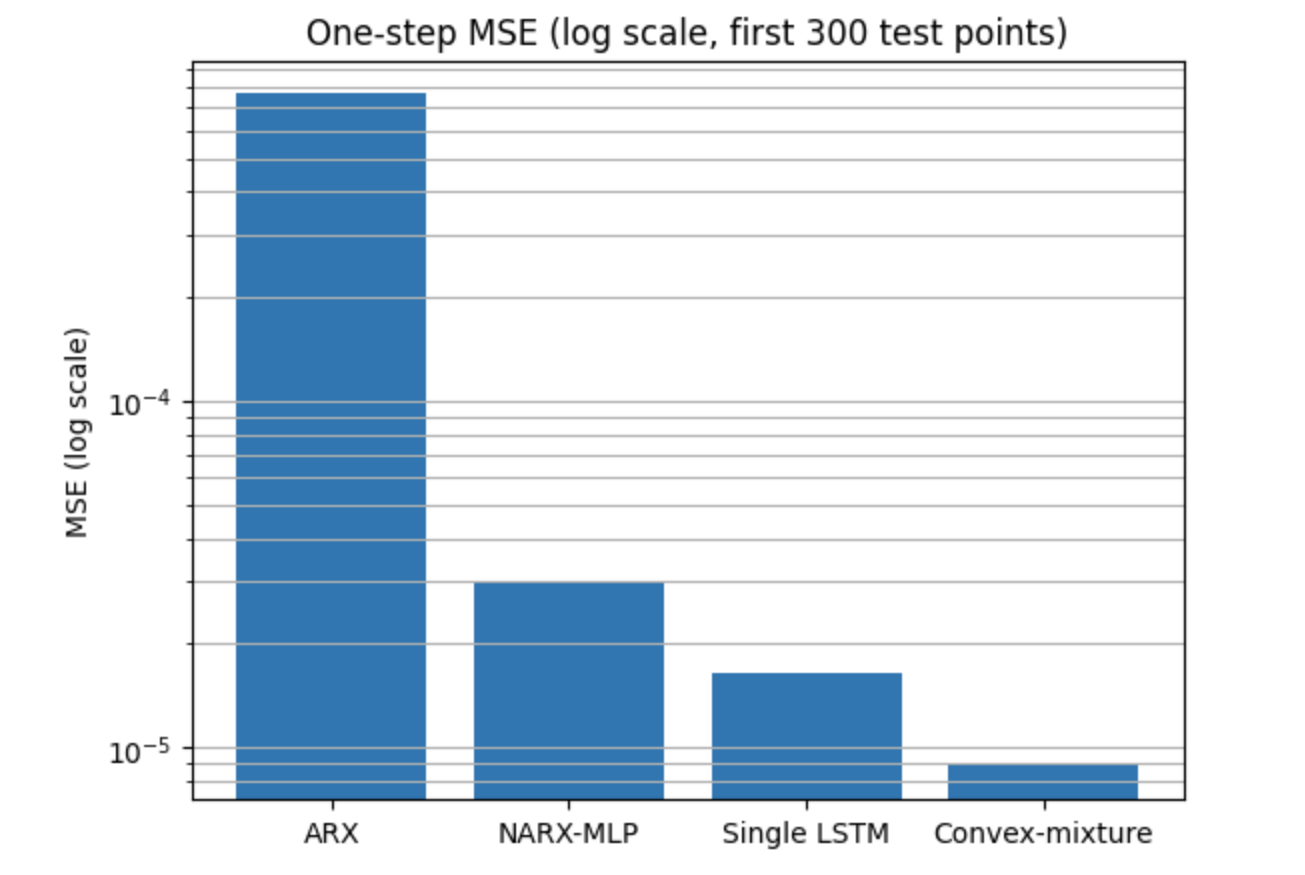}
  \vspace{-2pt}
  \caption{One-step MSE (log scale).}
\end{subfigure}

\vspace{2pt}

\begin{subfigure}{0.9\linewidth}
  \centering
  \includegraphics[width=\linewidth]{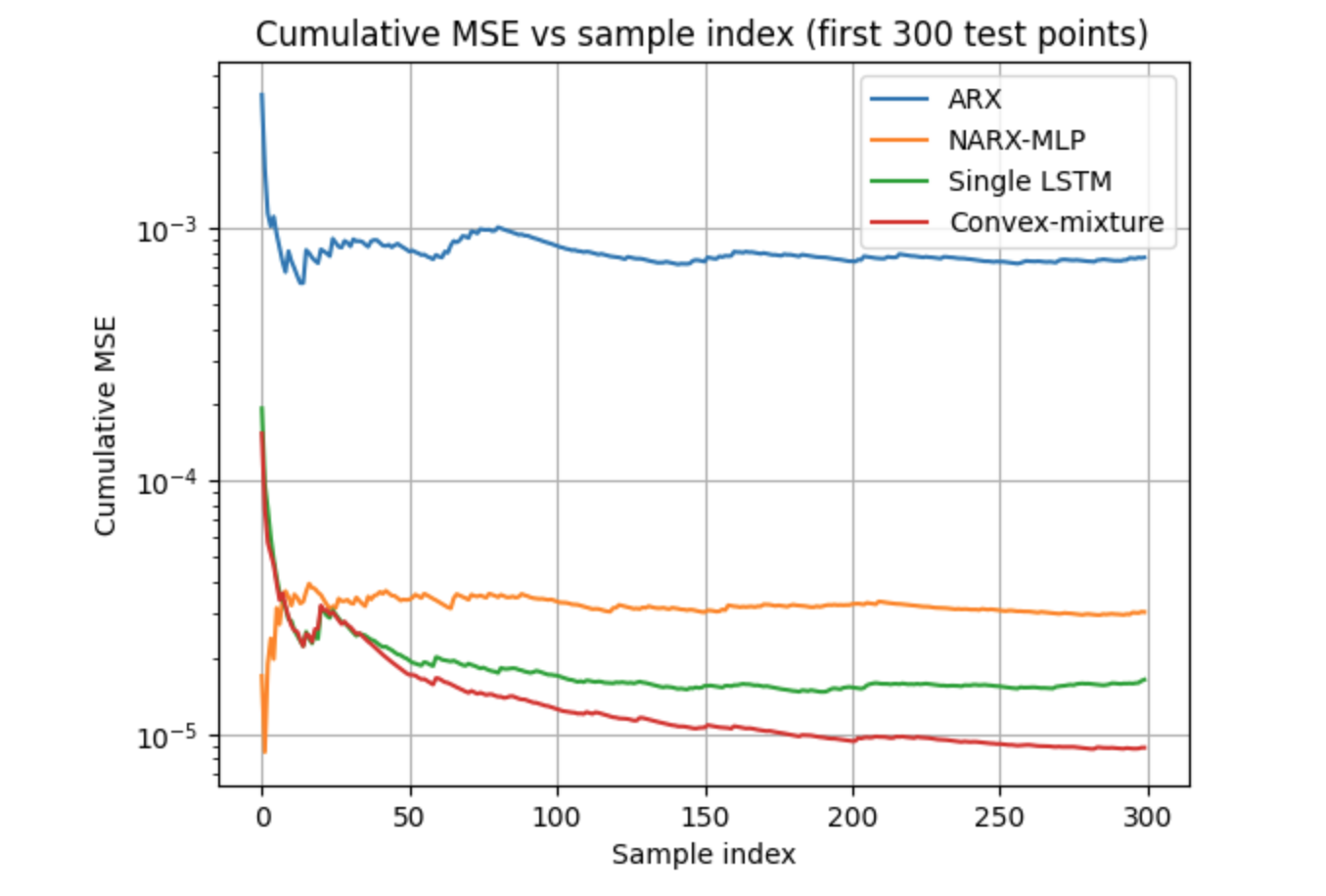}
  \vspace{-2pt}
  \caption{Cumulative MSE versus sample index.}
\end{subfigure}
\caption{Stationary NARMA-10 summary plots (first 300 test points).}
\label{fig:stationary}
\end{figure}

Table~\ref{tab:stationary} and Fig.~\ref{fig:stationary} summarize one-step MSE on a representative
test segment (first 300 points) for the baseline suite and the adaptive
convex mixture. As expected, ARX underfits the nonlinear recursion, while
NARX--MLP and LSTM reduce error by modeling nonlinearities and temporal
structure. The convex mixture provides additional improvement by
combining multiple experts.

\begin{table}[!h]
\renewcommand{\arraystretch}{1.2}
\caption{Stationary NARMA-10 One-Step MSE (First 300 Test Points).}
\vspace{-4pt}
\label{tab:stationary}
\centering
\begin{tabular}{@{}lc@{}}
\toprule
\textbf{Method} & \textbf{MSE} \\
\midrule
ARX                                     & $7.653\times 10^{-4}$ \\
NARX--MLP                               & $3.034\times 10^{-5}$ \\
Single LSTM                             & $1.640\times 10^{-5}$ \\
Adaptive convex mixture (3 experts)     & $8.833\times 10^{-6}$ \\
\bottomrule
\end{tabular}
\end{table}

\subsection{Regime Switching: Experts Trained on Full Switching Data}
\begin{figure}[!h]
\centering
\begin{subfigure}{0.98\linewidth}
  \centering
  \includegraphics[width=\linewidth]{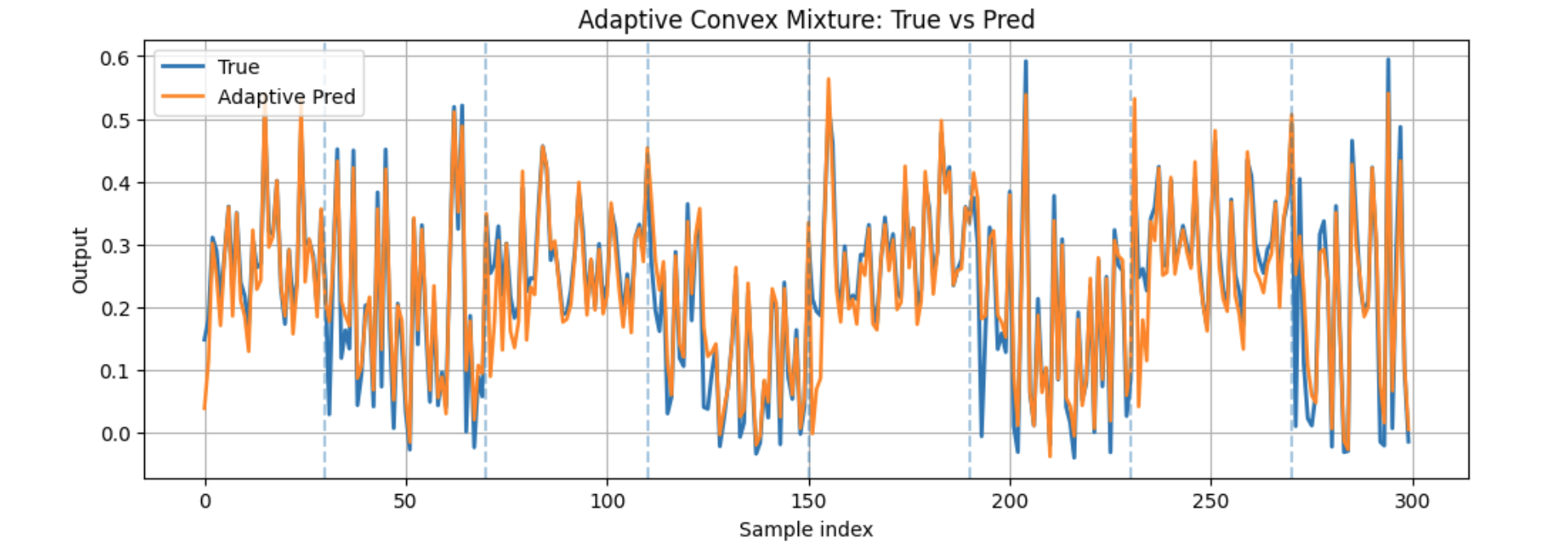}
  \caption{Adaptive update: True vs Pred.}
\end{subfigure}

\vspace{2pt}

\begin{subfigure}{0.98\linewidth}
  \centering
  \includegraphics[width=\linewidth]{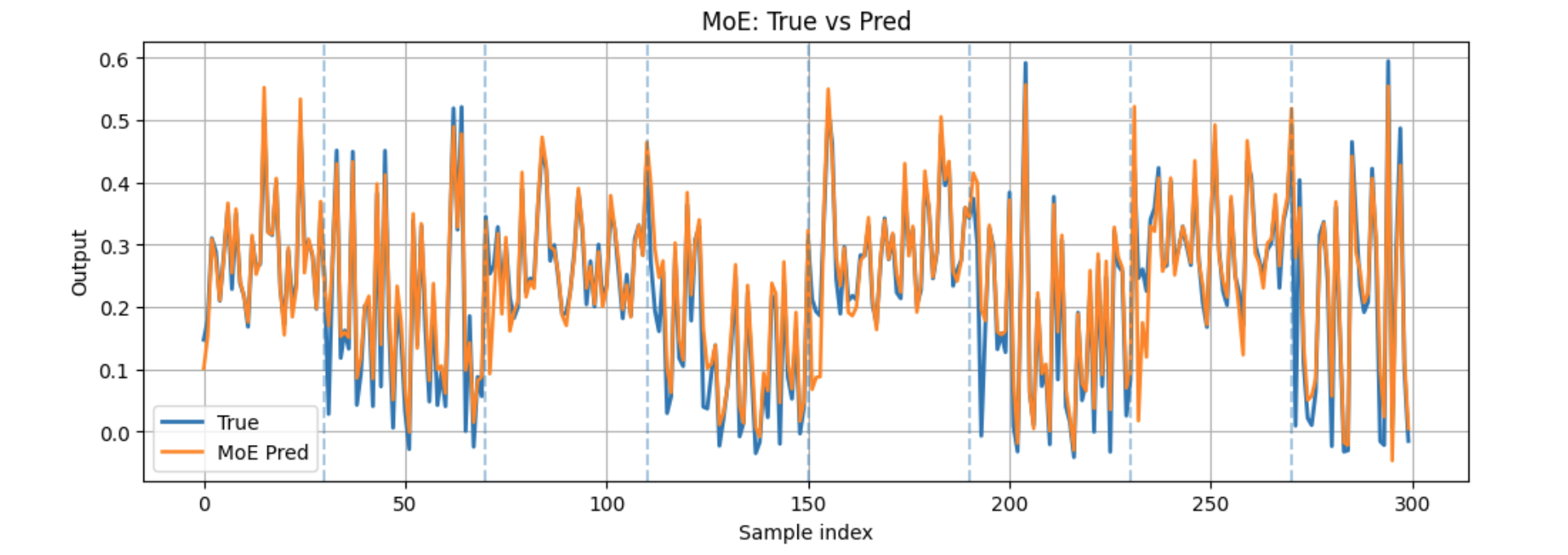}
  \caption{MoE gate: True vs Pred.}
\end{subfigure}
\caption{Regime-switching (full-data training): qualitative predictions
on a representative test segment.}
\label{fig:fulldata-qual}
\end{figure}

\begin{figure}[!h]
\centering
\begin{subfigure}{\linewidth}
  \centering
  \includegraphics[width=\linewidth]{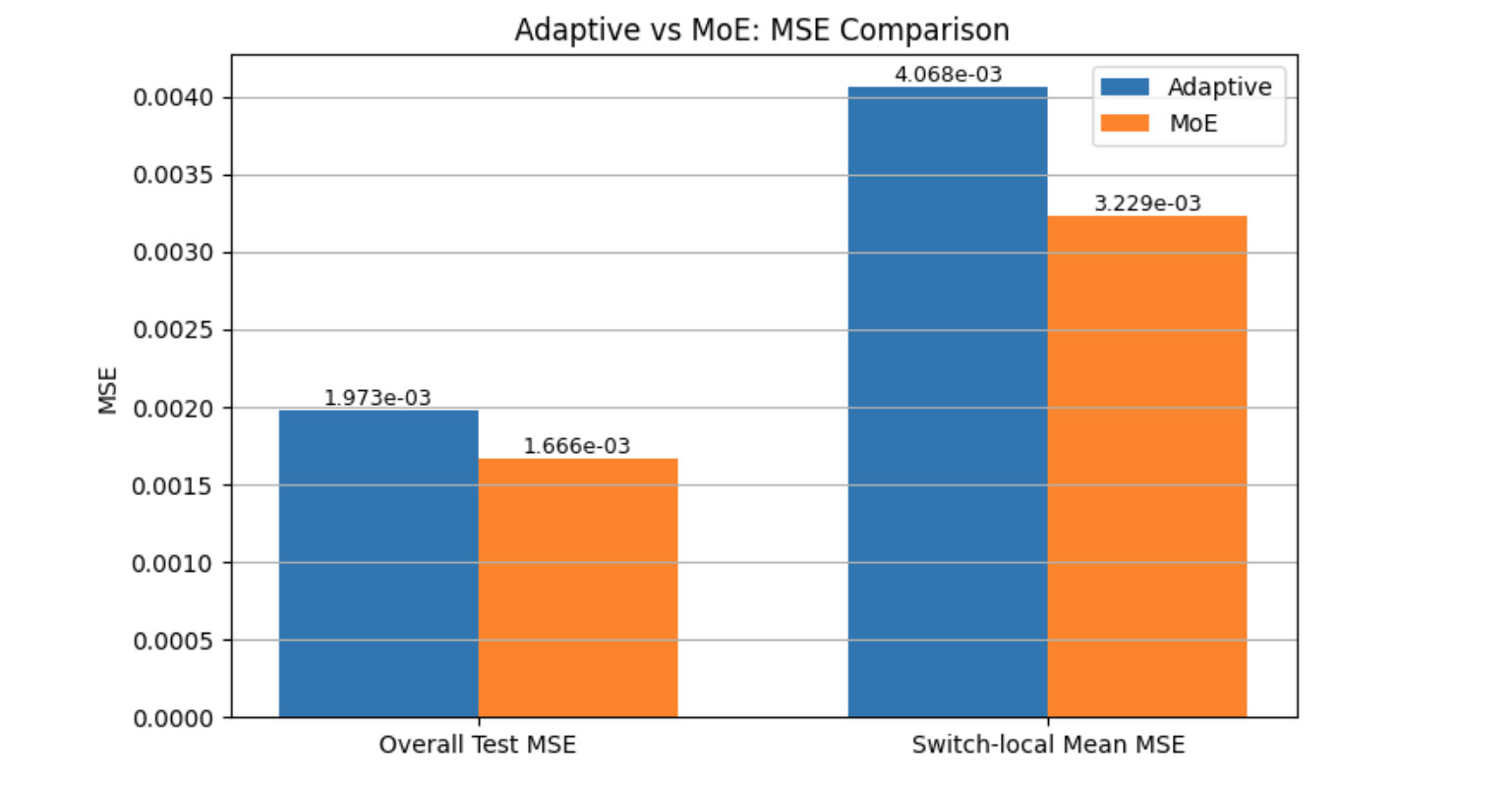}
  \caption{Overall and after-switch MSE.}
\end{subfigure}

\vspace{2pt}

\begin{subfigure}{\linewidth}
  \centering
  \includegraphics[width=\linewidth]{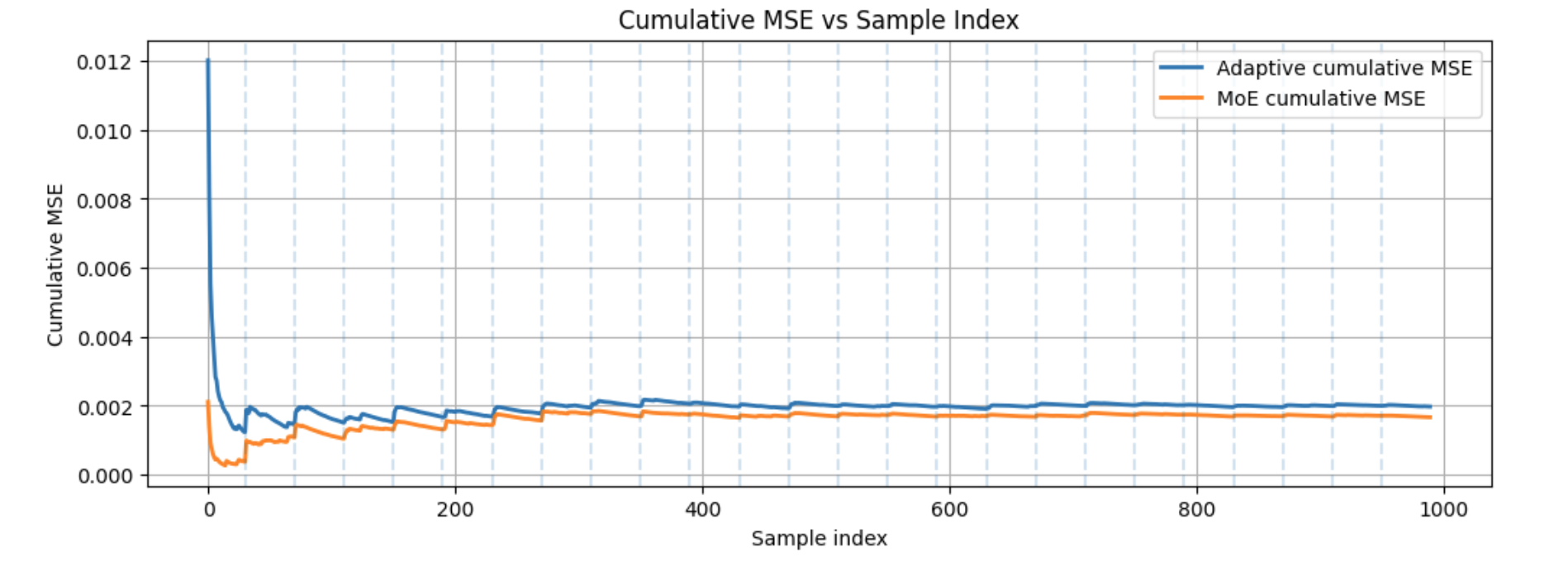}
  \caption{Cumulative MSE.}
\end{subfigure}
\caption{Regime-switching performance with experts trained on the full
switching dataset.}
\label{fig:fulldata-quant}
\end{figure}

We first compare adaptive convex mixture of experts and MoE gating network when all model components are trained on the full switching dataset. Fig.~\ref{fig:fulldata-qual}, Fig.~\ref{fig:fulldata-quant}, and Table~\ref{tab:fulldata} reports the overall test MSE ($\mathrm{MSE}_{\mathrm{all}}$) and after-switch window MSE ($\mathrm{MSE}_{\mathrm{sw}}$) defined in~\eqref{eq:mse-all} and \eqref{eq:mse-sw}. MoE shows a modest improvement in both metrics, but the gap is limited because experts trained on mixed-regime data tend to become generalists.

\begin{table}[!h]
\renewcommand{\arraystretch}{1.2}
\caption{Regime-switching NARMA-10 (full-data training): overall and after-switch MSE.}
\vspace{-4pt}
\label{tab:fulldata}
\centering
\setlength{\tabcolsep}{4pt}
\footnotesize
\begin{tabular}{@{}lcc@{}}
\toprule
\textbf{Method} & \textbf{Overall MSE} & \textbf{After-switch MSE} \\
\midrule
Adaptive convex mixture & $1.973\times 10^{-3}$ & $4.068\times 10^{-3}$ \\
MoE (WindowGate)        & $1.666\times 10^{-3}$ & $3.229\times 10^{-3}$ \\
\bottomrule
\end{tabular}
\end{table}

\subsection{Regime Switching: Frozen-Experts Ablation}

To isolate the mixing weights update rule, experts are pre-trained to specialize on
different regimes and then frozen. The test stream contains all regimes,
but expert outputs are identical across adaptive and MoE methods; only
the mixing weights update mechanism differs. Table~\ref{tab:frozen} shows that MoE
gating substantially reduces both overall and after-switch errors
compared to the adaptive update. Fig.~\ref{fig:frozen-qual} shows that the MoE gate tracks the true output
more closely after regime changes when the experts are frozen and
specialized. Fig.~\ref{fig:frozen-quant} and Table~\ref{tab:frozen}
quantify this behavior, showing that the learned gate substantially reduces
both overall and after-switch MSE relative to the adaptive update rule. The results support the main claim that is when specialization exists, a learned context-aware MoE can switch experts more effectively than a MoE with reactive error-driven rule.

\subsection{Context-aware MoE's advantage over adaptive mixture}
\begin{figure}[!h]   
\centering
\begin{subfigure}{\linewidth}
  \centering
  \includegraphics[width=\linewidth]{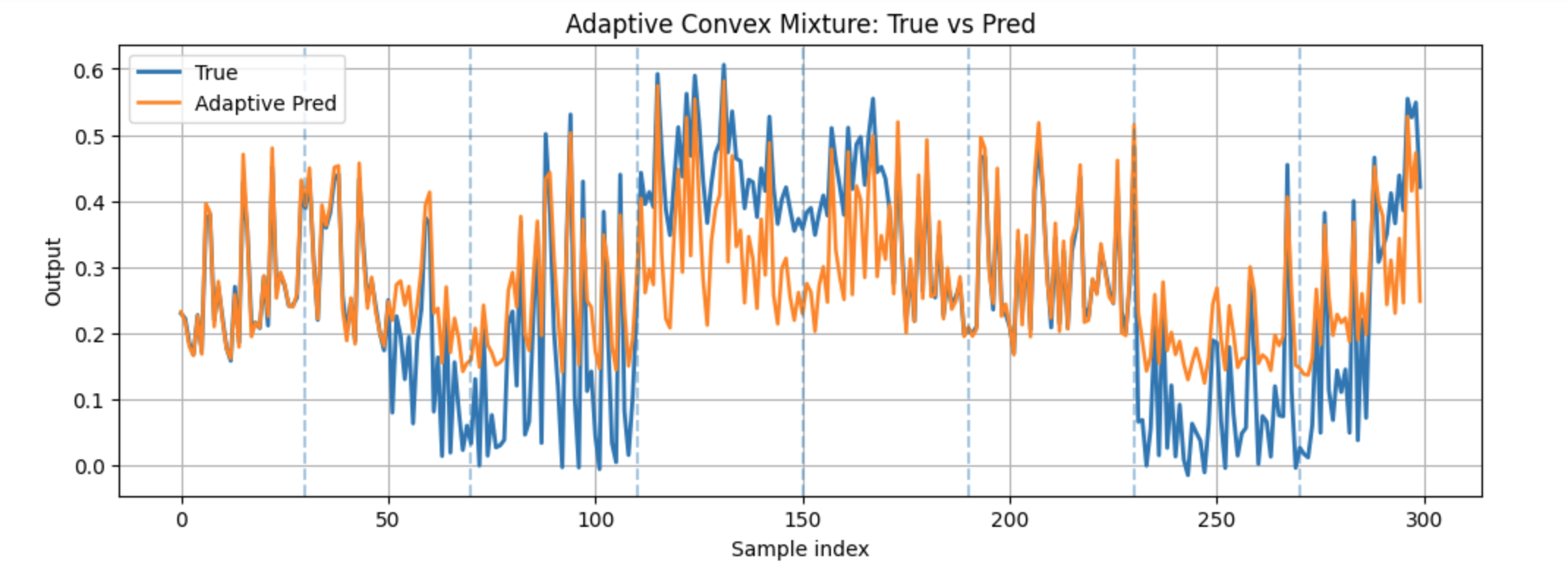}
  \caption{Adaptive update: True vs Pred.}
\end{subfigure}

\vspace{2pt}

\begin{subfigure}{\linewidth}
  \centering
  \includegraphics[width=\linewidth]{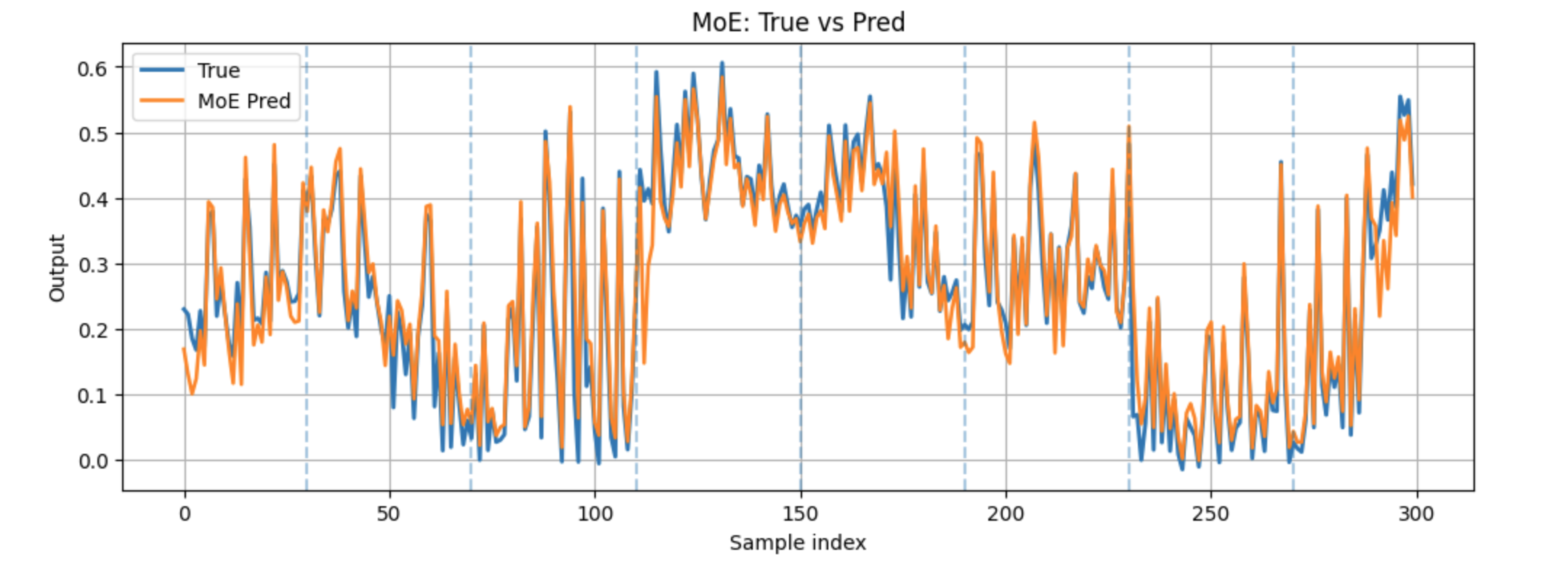}
  \caption{MoE gate: True vs Pred.}
\end{subfigure}
\caption{Frozen-experts ablation: qualitative predictions under regime switching.}
\label{fig:frozen-qual}
\end{figure}

\begin{figure}[!h]   
\centering
\begin{subfigure}{0.95\linewidth}
  \centering
  \includegraphics[width=\linewidth]{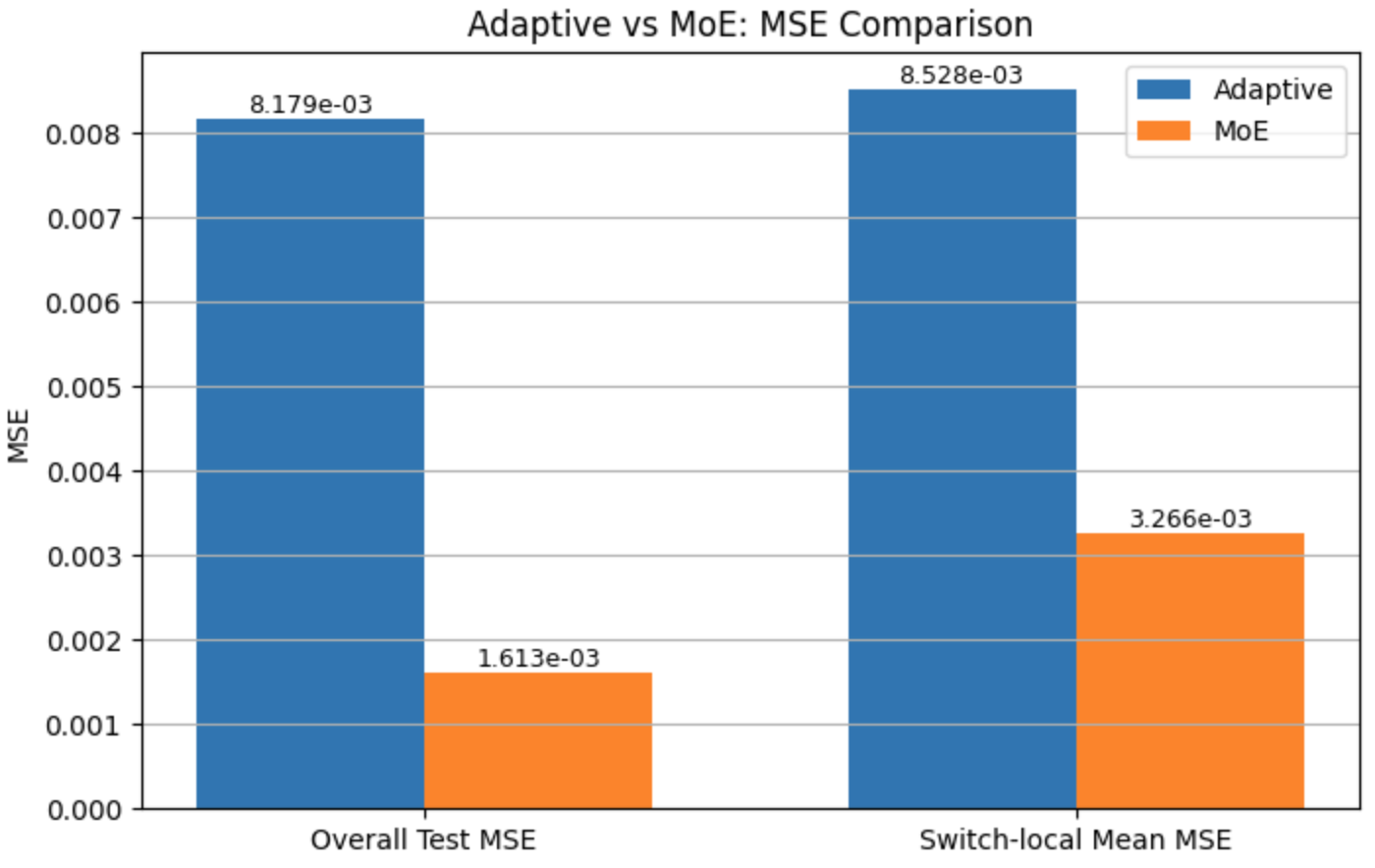}
  \caption{Overall and after-switch MSE.}
\end{subfigure}

\vspace{6pt}

\begin{subfigure}{0.98\linewidth}
  \centering
  \includegraphics[width=\linewidth]{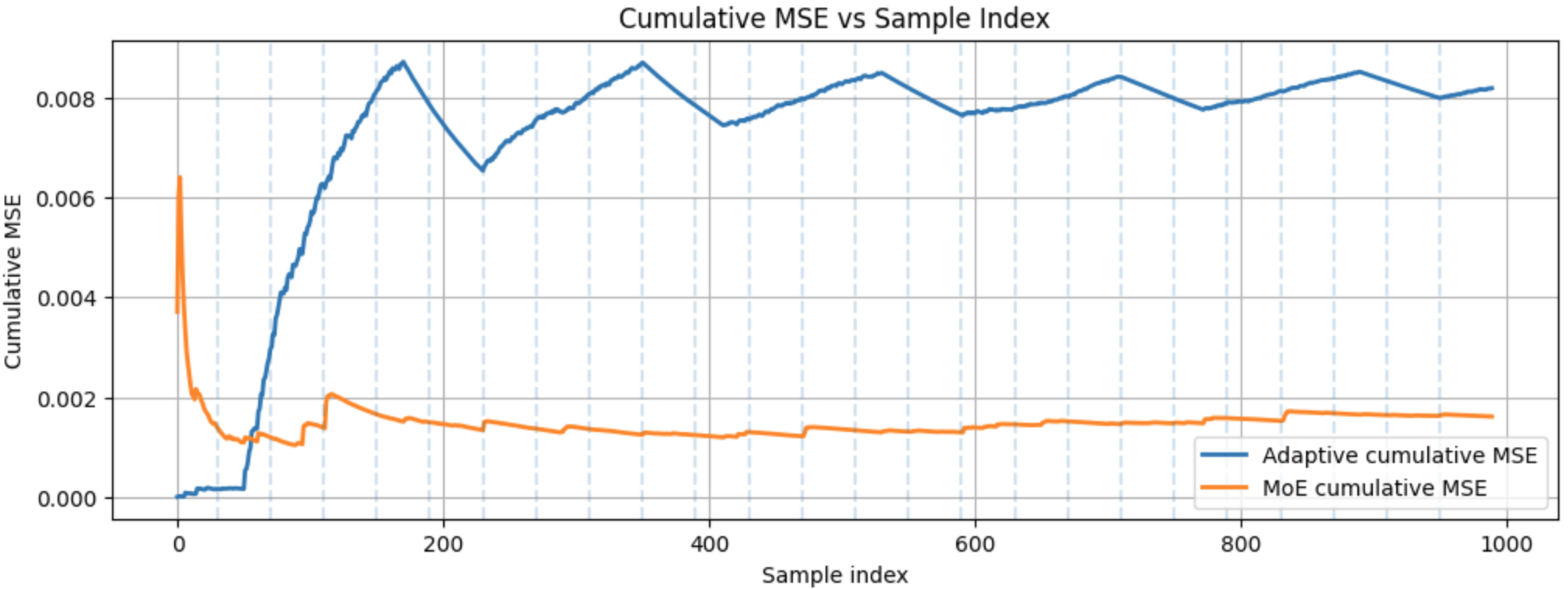}
  \caption{Cumulative MSE.}
\end{subfigure}
\caption{Regime-switching performance in the frozen-experts ablation.}
\label{fig:frozen-quant}
\end{figure}

The full-data switching experiment mixes regimes during training, so each
expert receives gradients from all operating conditions. In that setting,
experts tend to become generalists, and both mixing weights update rules (adaptive
update and MoE gating) converge toward using similar convex combinations;
therefore, gains from a smarter gate are modest. The frozen-experts
ablation removes this confounding factor; experts are specialized and fixed, so performance depends primarily on how quickly and accurately the mixing weights update mechanism selects the appropriate expert after a regime change. This is
the regime in which end-to-end gating is most beneficial, and the
results in Table~\ref{tab:frozen}, Fig.~\ref{fig:frozen-qual}, and Fig.~\ref{fig:frozen-quant} show a
large reduction in both $\mathrm{MSE}_{\mathrm{all}}$ and
$\mathrm{MSE}_{\mathrm{sw}}$.

\begin{table}[!h]
\renewcommand{\arraystretch}{1.2}
\caption{Frozen experts ablation (regime switching): MoE advantage when only the mixing weights update rule differs.}
\label{tab:frozen}
\centering
\setlength{\tabcolsep}{4pt}
\footnotesize
\begin{tabular}{@{}lcc@{}}
\toprule
\textbf{Method} & \textbf{Overall MSE} & \textbf{After-switch MSE} \\
\midrule
Adaptive (frozen)        & $8.177\times 10^{-3}$ & $8.528\times 10^{-3}$ \\
MoE gate (frozen)        & $1.613\times 10^{-3}$ & $3.266\times 10^{-3}$ \\
Factor (Adaptive/MoE)    & $5.07\times$          & $2.61\times$ \\
\bottomrule
\end{tabular}
\end{table}

\FloatBarrier

\section{Conclusion}
This paper studied convex mixtures of LSTM experts for nonlinear system
identification on NARMA-10 and compared two mixing weights update mechanisms: an
error-driven adaptive update with simplex projection and a learned MoE
gate. The proposed MoE formulation is end-to-end differentiable, removes
heuristic projection-based tuning, and can learn context-aware expert
selection patterns. On stationary NARMA-10, MoE achieves accuracy
comparable to the adaptive mixture, indicating that end-to-end gating can
match a strong reactive rule when the dynamics are unchanged in a single regime. On
regime-switching data, a frozen-experts ablation isolates the mixing weights update rule
and shows that MoE gating network approach provides a clear advantage when experts are
specialized and trained in a context-aware way, substantially reducing both overall and after-switch MSE errors.
Future work includes scaling to more regimes and experts, exploring
sparse top-$k$ gating, and extending evaluation to multi-step rollouts
where error accumulation becomes critical. Another direction is online
adaptation under measurement noise, where the gate must remain stable
while tracking gradual drift.

	\bibliographystyle{IEEEtran}
	\bibliography{references}

@incollection{ljung1998system,
  title={System identification},
  author={Ljung, Lennart},
  booktitle={Signal analysis and prediction},
  pages={163--173},
  year={1998},
  publisher={Springer}
}

@book{billings2013nonlinear,
  title={Nonlinear system identification: NARMAX methods in the time, frequency, and spatio-temporal domains},
  author={Billings, Stephen A},
  year={2013},
  publisher={John Wiley \& Sons}
}

@article{lin1996learning,
  title={Learning long-term dependencies in NARX recurrent neural networks},
  author={Lin, Tsungnan and Horne, Bill G and Tino, Peter and Giles, C Lee},
  journal={IEEE transactions on neural networks},
  volume={7},
  number={6},
  pages={1329--1338},
  year={1996},
  publisher={IEEE}
}

@article{hochreiter1997long,
  title={Long short-term memory},
  author={Hochreiter, Sepp and Schmidhuber, J{\"u}rgen},
  journal={Neural computation},
  volume={9},
  number={8},
  pages={1735--1780},
  year={1997},
  publisher={MIT press}
}

@article{srivastava2016lstm,
  title={LSTM: A search space odyssey},
  author={Srivastava, Rupesh},
  journal={IEEE transactions on neural networks and learning systems},
  year={2016}
}

@article{jacobs1991adaptive,
  title={Adaptive mixtures of local experts},
  author={Jacobs, Robert A and Jordan, Michael I and Nowlan, Steven J and Hinton, Geoffrey E},
  journal={Neural computation},
  volume={3},
  number={1},
  pages={79--87},
  year={1991},
  publisher={MIT Press}
}

@article{jordan1994hierarchical,
  title={Hierarchical mixtures of experts and the EM algorithm},
  author={Jordan, Michael I and Jacobs, Robert A},
  journal={Neural computation},
  volume={6},
  number={2},
  pages={181--214},
  year={1994},
  publisher={MIT Press}
}

@article{shazeer2017outrageously,
  title={Outrageously large neural networks: The sparsely-gated mixture-of-experts layer},
  author={Shazeer, Noam and Mirhoseini, Azalia and Maziarz, Krzysztof and Davis, Andy and Le, Quoc and Hinton, Geoffrey and Dean, Jeff},
  journal={arXiv preprint arXiv:1701.06538},
  year={2017}
}

@article{yuksel2012twenty,
  title={Twenty years of mixture of experts},
  author={Yuksel, Seniha Esen and Wilson, Joseph N and Gader, Paul D},
  journal={IEEE transactions on neural networks and learning systems},
  volume={23},
  number={8},
  pages={1177--1193},
  year={2012},
  publisher={IEEE}
}

@article{hamilton1989new,
  title={A new approach to the economic analysis of nonstationary time series and the business cycle},
  author={Hamilton, James D},
  journal={Econometrica: Journal of the econometric society},
  pages={357--384},
  year={1989},
  publisher={JSTOR}
}

@book{do2005discrete,
  title={Discrete-time Markov jump linear systems},
  author={Do Costa, Oswaldo Luiz Valle and Marques, Ricardo Paulino and Fragoso, Marcelo Dutra},
  year={2005},
  publisher={Springer}
}

@article{amani2026learning,
  title={Learning optimal crew dispatch for grid restoration following an earthquake},
  author={Amani, Farshad and Ardali, Faezeh and Kargarian, Amin},
  journal={IEEE Transactions on Smart Grid},
  year={2026},
  publisher={IEEE}
}

@article{amani2025learning,
  title={Learning Interior Point Method for AC and DC Optimal Power Flow},
  author={Amani, Farshad and Kargarian, Amin and Vaidyanathan, Ramachandran},
  journal={arXiv preprint arXiv:2508.19146},
  year={2025}
}

@inproceedings{wang2017new,
  title={A new concept using LSTM neural networks for dynamic system identification},
  author={Wang, Yu},
  booktitle={2017 American control conference (ACC)},
  pages={5324--5329},
  year={2017},
  organization={IEEE}
}

@article{atiya2000new,
  title={New results on recurrent network training: unifying the algorithms and accelerating convergence},
  author={Atiya, Amir F and Parlos, Alexander G},
  journal={IEEE transactions on neural networks},
  volume={11},
  number={3},
  pages={697--709},
  year={2000},
  publisher={IEEE}
}

@inproceedings{duchi2008efficient,
  title={Efficient projections onto the l 1-ball for learning in high dimensions},
  author={Duchi, John and Shalev-Shwartz, Shai and Singer, Yoram and Chandra, Tushar},
  booktitle={Proceedings of the 25th international conference on Machine learning},
  pages={272--279},
  year={2008}
}

@article{kingma2014adam,
  title={Adam: A method for stochastic optimization},
  author={Kingma, Diederik P and Ba, Jimmy},
  journal={arXiv preprint arXiv:1412.6980},
  year={2014}
}

@article{yin2023sequence,
  title={Sequence-to-sequence LSTM-based dynamic system identification of piezo-electric actuators},
  author={Yin, Ruocheng and Ren, Juan},
  year={2023},
  publisher={IEEE}
}

@inproceedings{bongiovanni2024data,
  title={Data-driven nonlinear system identification of a throttle valve using Koopman representation},
  author={Bongiovanni, Nicolas and Mavkov, Bojan and Martins, Renato and Allibert, Guillaume},
  booktitle={2024 American Control Conference (ACC)},
  pages={92--97},
  year={2024},
  organization={IEEE}
}

@article{revay2020convex,
  title={A convex parameterization of robust recurrent neural networks},
  author={Revay, Max and Wang, Ruigang and Manchester, Ian R},
  journal={IEEE Control Systems Letters},
  volume={5},
  number={4},
  pages={1363--1368},
  year={2020},
  publisher={IEEE}
}

@article{paoletti2007identification,
  title={Identification of hybrid systems a tutorial},
  author={Paoletti, Simone and Juloski, Aleksandar Lj and Ferrari-Trecate, Giancarlo and Vidal, Ren{\'e}},
  journal={European journal of control},
  volume={13},
  number={2-3},
  pages={242--260},
  year={2007},
  publisher={Elsevier}
}

@inproceedings{ahmed2011variational,
  title={Variational learning of autoregressive mixtures of experts for fully Bayesian hybrid system identification},
  author={Ahmed, Nisar and Campbell, Mark},
  booktitle={Proceedings of the 2011 American Control Conference},
  pages={139--144},
  year={2011},
  organization={IEEE}
}

@article{amani2026event,
  title={Event-Driven Deep RL Dispatcher for Post-Storm Distribution System Restoration},
  author={Amani, Farshad and Ardali, Faezeh and Kargarian, Amin},
  journal={arXiv preprint arXiv:2601.10044},
  year={2026}
}
\end{document}